\documentclass[sigplan,10pt]{acmart}
\AtBeginDocument{%
  }

\setcopyright{none}
\copyrightyear{}
\acmYear{}
\acmDOI{}
\acmConference{}{}{}
\acmISBN{}

\renewcommand\footnotetextcopyrightpermission[1]{}
\makeatletter
\renewcommand\acmConference[3][]{}
\makeatother

\usepackage{mathpartir}
\usepackage{stmaryrd}
\usepackage{todonotes}
\usepackage{xspace}
\usepackage{cleveref}
\usepackage{listings}

\usepackage[binary-units=true,detect-weight=true]{siunitx}

\newcommand{\vampire}{\textsc{Vampire}\xspace}
\newcommand{\avatar}{\textsc{Avatar}\xspace}
\newcommand{\cadical}{CaDiCaL\xspace}
\newcommand{\minisat}{\textsc{MiniSat}\xspace}
\newcommand{\dedukti}{\textsc{Dedukti}\xspace}
\newcommand{\lambdapi}{\textsc{lambdapi}\xspace}
\newcommand{\false}{\bot}
\newcommand{\prf}[1]{\mathsf{Prf}~#1}
\newcommand{\El}[1]{\mathsf{El}~#1}
\newcommand{\eq}[2]{(\mathsf{eq}~#1~#2)}
\newcommand{\deep}[1]{|#1|}
\newcommand{\shallow}[1]{\llbracket #1 \rrbracket}
\newcommand{\prop}{\mathsf{Prop}}
\newcommand{\inhabit}[1]{\star_#1}
\newcommand{\spl}[1]{\mathsf{sp}_{#1}}
\newcommand{\mgu}{\mathsf{mgu}}

\newcommand{\bnfdef}{::=}      % definition symbol
\newcommand{\bnfalt}{\;|\;}    % alternative symbol
\newcommand{\bnf}[1]{\textit{#1}}   % nonterminal (italic)
\begin{document}

\title[Verified \vampire Proofs]{Verified \vampire Proofs\texorpdfstring{\\}{}in the \texorpdfstring{$\lambda\Pi$}{λΠ}-Calculus Modulo Theories}

\author{Anja Petkovi\'c Komel}
\email{anja@argot.org}
\orcid{0000-0001-7203-6641}
\affiliation{%
  \institution{Argot Collective}
  \city{Zug}
  \country{Switzerland}
}

\author{Michael Rawson}
\affiliation{%
  \institution{University of Southampton}
  \city{Southampton}
  \country{United Kingdom}}
\email{michael@rawsons.uk}
\orcid{0000-0001-7834-1567}

\author{Martin~Suda}
\email{martin.suda@cvut.cz}
\orcid{0000-0003-0989-5800}
\affiliation{%
  \institution{Czech Technical University in Prague}
  \city{Prague}
  \country{Czech Republic}
}

% \renewcommand{\shortauthors}{Komel et al.}

\begin{abstract}
    The \vampire automated theorem prover is extended to output machine-checkable proofs in the \dedukti concrete syntax for the $\lambda\Pi$-calculus modulo.
    This significantly reduces the trusted computing base and in principle eases proof reconstruction in other proof-checking systems.
    Existing theory is adapted to deal with \vampire's internal logic and inference system.
    Implementation experience is reported, encouraging adoption of verified proofs in other automated systems.
\end{abstract}




% \received{5 September 2025}

\maketitle

\section{Introduction}

Theorem provers are pieces of software that help a user specify and prove mathematical statements, including properties of computer systems.
There are two kinds of theorem provers: automatic theorem provers (ATPs) are fully automatic, the user just provides the axioms and the conjecture and lets the prover find a proof, and interactive theorem provers (ITPs) or proof assistants that guide and assist the user towards constructing a proof while checking validity of all deduction steps.

ATPs are now advanced, and therefore complex, pieces of software.
In order to trust that the proofs they produce are indeed proofs, we would like to have an external piece of software check the proofs.
The checker would preferably have a small trusted \emph{kernel} so that the trusted computing base is minimised.
We report our experience adapting the state-of-the-art \vampire ATP to emit \dedukti (ITP) proofs.
To our knowledge this is the first time this has been attempted in a system like \vampire.
While in principle it is known how to do this, in our project we found there were several missing pieces.
After laying out some necessary background and carefully defining the scope of the project, we describe some differences between theory and practice in Section~\ref{sec:bridging}, handle \vampire's \avatar architecture in Section~\ref{sec:avatar}, and successfully check many \vampire proofs in Section~\ref{sec:experiments}, some very large.

\section{Background}
\vampire~\cite{DBLP:conf/cav/BartekBCHHHKRRRSSV25} is a successful first-order ATP extended to reason with theories, induction, higher-order logic, and more.
\vampire employs a variety of techniques to find proofs.
It is a highly complex piece of software with a large trusted code base: bugs are regularly found~\cite{CASC-J10,DBLP:conf/tap/Reger0V17} and some are likely still hidden.
If \vampire finds a proof, it emits a directed graph of formulae, together with information about what inference rule was used to deduce a formula from which premises.
In principle this proof should provide enough evidence~\cite{TSTP} for a human to reconstruct the full \emph{proof} with all the details, but in practice \vampire proofs can be long and involved, so manually checking them may not be desirable or feasible.

\dedukti~\cite{dedukti} is designed to efficiently machine-check large proofs while being interoperable with interactive theorem provers (ITPs). The underlying $\lambda \Pi$-calculus modulo theory is the proposed standard for proof interoperability~\cite{Boespflug2012TheM,DowekEmbeddingPTS}.
It is an expressive enough logical framework to enable shallow embedding of many theories, meaning the $\lambda \Pi$ constructors, representation of binders, and rewrite rules can be used to directly express derivations of other theories, rather than explicitly specifying the grammar and its interpretation. It has been shown~\cite{DowekEmbeddingPTS} that we can shallowly embed into $\lambda\Pi$ higher-order logics~\cite{assaf2015:HOLDedukti} (the underlying theory of proof assistants including Isabelle/HOL~\cite{Isabelle-HOL}), pure type systems such as Calculus of Inductive Constructions~\cite{Boespflug2012CoqInETT} (the underlying type theory of proof assistants inluding Coq~\cite{coq} and Lean~\cite{lean}), Martin-Löf type theory~\cite{GenestierAgdaToLambdapi2020,felicissimo2021:AgdaDedukti} (the underlying type theory of the proof assistant Agda~\cite{agda}), and the theories of other proof assistants (like PVS~\cite{hondet2020:PVSLambdapi} and Matita~\cite{assaf2015:MatitaDedukti}). There is also a proposed standard for shallow embedding of first-order logic in  $\lambda \Pi$ Calculus Modulo Theories, serving interoperability of first-order proofs.
Some ATPs, such as Zenon Modulo~\cite{zenon-modulo,burel2020firstorder}, iProver Modulo~\cite{burel,burel2020firstorder}, ArchSat~\cite{Bury2018SMTSM}, and Leo~III~\cite{taprogge-melanie-master-leo} can already emit proofs in this \dedukti/\lambdapi-checkable format.

\subsection{The $\lambda\Pi$-calculus modulo theories}

The $\lambda\Pi$-calculus modulo theories~\cite{lambdapi-calculus-modulo1,lambdapi-calculus-modulo2} is an extension of the $\lambda\Pi$-calculus, a proof language for minimal first-order logic based on the Curry-Howard-DeBruijn correspondence. Proofs are represented by $\lambda$-terms and formulas by their types, and it can be seen a Pure Type System.
We follow the presentation used by Burel~\cite{burel}.

Pre-terms are given by the following grammar:
\[
\begin{array}{rcl}
\bnf{M}, \bnf{N}, \bnf{A}, \bnf{B}
  & \bnfdef &
    \bnf{x} \bnfalt \bnf{$\lambda$ x : A . M} \bnfalt \bnf{$\Pi$ x : A . B} \\[4pt]
  &         & \bnfalt \bnf{M} \; \bnf{N}
              \bnfalt \bnf{$\mathsf{Type}$}
              \bnfalt \bnf{$\mathsf{Kind}$}
\end{array}
\]
where $x$ is an element of an infinite set of variables. A context is a set of couples of variables and pre-terms. A pre-term is called a term when it is well-typed according to the rules presented in~\Cref{fig:lambdapitypes}. The judgement $\vdash \Gamma$ is read ``Context $\Gamma$ is well-formed'' and the judgement $\Gamma \vdash T : A$ means the term $T$ has type $A$ in the context $\Gamma$. The conversion rule only applies if the types $A$ and $B$ are $\beta$-equal, which we marked by $A \equiv_\beta B$. In $\lambda\Pi$-calculus \emph{modulo} theories, this conversion rule is extended with the well-typed rewriting rules.

\begin{figure}
	\begin{mathpar}
	\inferrule[Empty Context]
	{ }
	{\vdash \emptyset}
	\and
	\inferrule[Variable Declaration]
	{
		\vdash \Gamma \\ 
		\Gamma \vdash A : s \\
		x \notin \Gamma
	}
	{
		\vdash \Gamma, x:A
	}
	\and 
	\inferrule[Sort]
	{
		\vdash \Gamma 
	}
	{
		\Gamma \vdash \mathsf{Type} : \mathsf{Kind}
	}
	\and
	\inferrule[Variable]
	{
		\vdash \Gamma \\ 
		(x : A) \in \Gamma 
	}
	{
		\Gamma \vdash x : A
	}
	\and
	\inferrule[Product]
	{
		\Gamma \vdash A : \mathsf{Type} \\ 
		\Gamma, x : A \vdash B : s 
	}
	{
		\Gamma \vdash \Pi x : A.~B : s
	}
	\and
	\inferrule[Application]
	{
		\Gamma \vdash T : \Pi x : A.~B\\
		\Gamma \vdash U : A
	}
	{
		\Gamma \vdash T~U : B[u/x]
	}
	\and
	\inferrule[Lambda]
	{
		\Gamma \vdash A : \mathsf{Type}\\ 
		\Gamma, x : A \vdash B : s\\ 
		\Gamma, x : A \vdash T : B
	}
	{
		\Gamma \vdash \lambda x : A.~T : \Pi x : A.~B
	}
	\and
	\inferrule[Conversion]
	{
		\Gamma \vdash T : A \\ 
		\Gamma \vdash A : s \\ 
		\Gamma \vdash B : s
	}
	{
		\Gamma \vdash T : B
	}
	A \equiv_\beta B
	\end{mathpar}
	\caption{Typing rules for $\lambda\Pi$-calculus where the parameter $s~\in~\{\mathsf{Type}, \mathsf{Kind}\}$. The conversion rule only applies if the types are $\beta$-equal.}
	\label{fig:lambdapitypes}
\end{figure}

\begin{definition}[Rewrite rule]
	A \emph{rewrite rule} is a quadruple $(\Delta, l, r, A)$, where $\Delta$ is a context and $l$, $r$ and $A$ are terms. It is well-typed in a context $\Gamma$ if
	\begin{itemize}
		\item $\vdash \Gamma, \Delta$ is derivable, and 
		\item $\Gamma, \Delta \vdash l : A$ and $\Gamma, \Delta \vdash r : A$ are derivable.
	\end{itemize}
	We write $l \hookrightarrow r$ when context $\Delta$ and type $\Gamma$ can be inferred. 
\end{definition}

The context $\Delta$ in the definition of a rewrite rule plays a role of typing variables in the rule
and $A$ ensures that $l$ and $r$ have the same type, so that the types are preserved through rewriting. A term $t$ is rewritten by a rewrite rule using a suitable substitution of the variables in $\Delta$.

In the $\lambda\Pi$-calculus modulo theories, contexts can contain rewriting rules. The type system form~\Cref{fig:lambdapitypes} is extended by a rule that adds a well-typed rewriting rule to a context:
\begin{mathpar}
	\inferrule[Rewrite]
	{
		\Gamma, \Delta \vdash l : A \\ 
		\Gamma, \Delta \vdash r : A 
	}
	{
		\vdash \Gamma, (\Delta, l, r, A)
	}
\end{mathpar}
Given a context $\Gamma$, let $\equiv_\Gamma$ be the smallest congruence relation generated by $\beta$-reduction and the rewriting rules of $\Gamma$. The conversion rule of the $\lambda\Pi$-calculus is then replaced by the following one:
\begin{mathpar}
	\inferrule[Conversion]
	{
		\Gamma \vdash T : A \\ 
		\Gamma \vdash A : s \\ 
		\Gamma \vdash B : s
	}
	{
		\Gamma \vdash T : B
	}
	A \equiv_\Gamma B
\end{mathpar}
where $s \in \{\mathsf{Type}, \mathsf{Kind}\}$. A \emph{definition} is a special case of a rewrite rule where the context $\Delta$ is empty and $l$ only contains one symbol.

In \dedukti, the syntax for $\lambda x : A.~t$ and $\Pi x : A.~B$ are written \texttt{x : A => t} and \texttt{x : A -> B} respectively. A rewriting rule $(\Delta, l, r, A)$ is declared with \texttt{[$\Delta$] l --> r}. The \dedukti type checker \texttt{dk check} can check that a context presented by a \dedukti file is well-formed and in particular it checks that rewriting rules are well-typed. If in the context there is a declaration of a constant $c$ of type $A$ and a rule rewriting $c$ into a term $t$, the fact that the context is well-formed implies that $t$ has type $A$ and by the Curry-Howard correspondence this means that $t$ is a proof of $A$. Similarly, if a constant of type $B$ is declared, but it is not rewritten, this can be seen as assuming the axiom $B$.

An alternative and increasingly popular concrete syntax for the $\lambda\Pi$-calculus modulo theories is the \lambdapi proof assistant. While \dedukti is designed for machine-checking large proofs, \lambdapi offers capabilities that are necessary for manual proof development, like inspecting the proofs by stepping through them.
Since \vampire produces some quite large proofs, we chose \dedukti for efficiency. It should be rather straightforward to translate proofs from \dedukti to \lambdapi for better integration and interoperability, but some engineering effort is needed (and remains to be done).

\subsection{Encoding First-Order Logics in $\lambda\Pi$-modulo}
\label{sec:encoding}

% would avoid saying ``mathematical foundation'', this suggests Vampire _has_ such a foundation...
We use the standard encoding of first-order provability~\cite{burel,universeUpaper}, which embeds first-order logic with a type former $\prf{X}$ for ``the type of proofs of $X$'' and a translation $\deep{\_}$ from first-order objects to terms in $\lambda\Pi$-modulo. 

Specifically, as given in~\Cref{eq:encoding-sorts}, this encoding introduces a type of propositions $\prop$, and a type former $\prf{}$. We assume a single sort $\iota$ of individuals until Section~\ref{sec:polymorphism}, where this assumption is relaxed.
For this reason we also assume an explicit type $\mathsf{Set} : \mathsf{Type}$ of sorts (so $\iota : \mathsf{Set}$) and a decoding function $\mathsf{El}$ to denote the type of elements.

\begin{align}
	\prop &: \mathsf{Type} \nonumber \\ 
	\prf{} &: \prop \to \mathsf{Type} \nonumber \\ 
	\iota &: \mathsf{Set} \label{eq:encoding-sorts} \\ 
	\mathsf{Set} &: \mathsf{Type} \nonumber \\ 
	\mathsf{El} &: \mathsf{Set} \to \mathsf{Type} \nonumber
\end{align}

The logical connectives are encoded as constants with the following types:
\begin{align*}
	\false &: \prop \\ 
	\neg &: \prop \to \prop \\
	\vee &: \prop \to \prop \to \prop \\ 
	\wedge &: \prop \to \prop \to \prop \\ 
	\Rightarrow &: \prop \to \prop \to \prop \\ 
	\forall &: (\El{\iota} \to \prop) \to \prop \\ 
	\exists &: (\El{\iota} \to \prop) \to \prop \\ 
\end{align*}

For every function symbol $f$ of arity $n$ we add a constant $f : \underbrace{\El{\iota} \to \cdots \to \El{\iota}}_{n \text{-times}} \to \El{\iota}$ and similarly for every predicate $P$ of arity $m$ we add $P : \underbrace{\El{\iota} \to \cdots \to \El{\iota}}_{m \text{-times}} \to \prop$.

We use a Leibniz encoding of equality with type $\mathsf{eq} : \El{\iota} \to \El{\iota} \to \mathsf{Prop}$.
The translation $\deep{\_}$ from first-order objects to terms in $\lambda\Pi$-modulo is given inductively on the structure of formulas and terms as follows (using the same symbols on both sides of the equality $\triangleq$,
where on the left it refers to the first-order object and on the right to the constant in $\lambda\Pi$-modulo encoding):
\begin{align*}
    \deep{x} &\triangleq x\\
    \deep{f(t_1, \ldots, t_n)} &\triangleq f~\deep{t_1}~\ldots~\deep{t_n}\\
    \deep{P(t_1, \ldots, t_n)} &\triangleq (P~\deep{t_1}~\ldots~\deep{t_n}) \\
    \deep{s = t} &\triangleq \eq{\deep{s}}{\deep{t}} \\
    \deep{\lnot L} &\triangleq \lnot \deep{L} \\
    \deep{L_1 \lor L_2 } &\triangleq \lor~\deep{L_1}~\deep{L_2} \\
    \deep{L_1 \land L_2 } &\triangleq \land~\deep{L_1}~\deep{L_2} \\
    \deep{L_1 \Rightarrow L_2 } &\triangleq ~\Rightarrow~\deep{L_1}~\deep{L_2} \\
    \deep{\forall X.~A} &\triangleq \forall~(\lambda X : \El{\iota}.~\deep{A}) \\
    \deep{\exists X.~A} &\triangleq \exists~(\lambda X : \El{\iota}.~\deep{A}) \\
\end{align*}

What makes the encoding shallow are the rewrite rules on the type former $\prf{}$ that relate the first-order connectives with their counterparts in $\lambda\Pi$-calculus modulo theories, as follows:
\begin{align*}
    \prf{\false} &\hookrightarrow \Pi r : \prop.~\prf{r}\\
    \prf{(\lnot p)} &\hookrightarrow \prf{p} \to \prf{\false} \\ 
    \prf{(p \Rightarrow q)} &\hookrightarrow \prf{p} \to \prf{q} \\
    \prf{(p \land q)} &\hookrightarrow \Pi r : \prop.~(\prf{p} \to \prf{q} \to \prf{r}) \\
                      &\hphantom{\hookrightarrow}\to \prf{r} \\
    \prf{(p \lor q)} &\hookrightarrow (\prf{p} \to \prf{\false}) \to (\prf{q} \to \prf{\false}) \\
                     &\hphantom{\hookrightarrow} \to \prf{\false} \\
    \prf{(\forall~f)} &\hookrightarrow \Pi x : \El{\iota}.~\prf{p~x} \\
    \prf{(\exists~f)} &\hookrightarrow \Pi r : \prop.~(\Pi x : \El{\iota}.~\prf{p~x} \to \prf{r}) \\
                      &\hphantom{\hookrightarrow}\to \prf{r}
\end{align*}
Note that the encoding of the first-order logic according to~\cite{burel} uses a more general version of disjunction, namely 
\begin{align}
	\prf{(p \lor q)} \hookrightarrow & \Pi r : \prop.~(\prf{p} \to \prf{r}) \to  \label{eq:or-general}\\
	&(\prf{q} \to \prf{r}) \to \prf{r} \nonumber
\end{align}
and it can be proved that with this general form, the full encoding is sound and a conservative extension of intuitionistic first-order logic.
We instantiate~\Cref{eq:or-general} with $\false$ for $r$. We made this design choice because it improves the human-readability of the proof (especially with chaining disjunctions in clauses). This choice is clearly sound and so far no issues were encountered with proof checking that would prompt us to use the full general form~(\ref{eq:or-general}).

We write $\shallow{O}$ for $\prf{\deep{O}} \to \prf{\false}$ and represent a (first-order) \emph{clause} $\forall x_1 ... x_n.~L_1 \lor \ldots \lor L_n$ as
\begin{equation}
\label{eq:clause}
\prod_{x_1 \ldots x_n : \El{\iota}}~\shallow{L_1} \to \ldots \to \shallow{L_n} \to \prf{\false}
\end{equation}
where $L_1$, \ldots, $L_n$ are literals (atomic formulas or their negations).
As shown in~\Cref{sec:inhabit} we need the axiom that domains are inhabited, so the encoding contains an explicit ``inhabit'' term $\star$ indexed by sorts:
\begin{equation}
\label{eq:inhabit-def}
\star : \prod_{A : \mathsf{Set}} \mathsf{El} \; A
\end{equation}

\subsection{\vampire's logic and calculus}
\label{sec:vampire-calculus}
\vampire reads an input problem in either of the TPTP~\cite{TPTP} or SMT-LIB~\cite{smtlib} formats and negates the conjecture if present, then proceeds with clausification (unless the problem is already in clausal normal form), and finally begins proof search to derive $\false$.
The internal language of \vampire (disregarding extensions) is a polymorphic classical first-order logic.
We refer the reader to standard material on first-order logic~\cite{smullyan} and to the rank-1 polymorphic logic TFF1~\cite{tff1} that \vampire implements.
There is no central proof \emph{kernel}, unlike ITPs or some other ATPs like Twee~\cite{twee}, and so the system as a whole is, by design, untrusted.

To achieve sound and complete reasoning, \vampire implements a core \emph{superposition} calculus~\cite{superposition} consisting of a handful of inferences.
These inferences will usually make up the bulk of a \vampire proof.
\vampire also implements a very large number of other inferences, but these are usually either (i) special cases of inferences of the superposition calculus; (ii) relate to transforming the input into the internal normal form; or (iii) relate to \vampire's extensions.
We will focus on inferences that appear after translation to normal form when using \vampire in the default mode.
This means the core superposition inferences and special cases like \emph{demodulation} must be handled.
At first we present the traditional single-sorted first-order logic, then extend it to many-sorted and polymorphic logic in~\Cref{sec:polymorphism}.

\begin{figure}
	\begin{mathpar}
		\inferrule[Resolution]
		{P \lor C \\ \neg Q \lor D}
		{\sigma(C \lor D)}
		\sigma = \mgu(P,Q)
		\and 
		\inferrule[Factoring]
		{
			L \lor K \lor C
		}
		{
			\sigma(L \lor C)
		}
		\sigma = \mgu(L, K)
		\and
		\inferrule[Paramodulation]
		{l = r \lor C \\ L[s] \lor E}
		{\sigma(L[r] \lor C \lor E)}
		\sigma = \mgu(l,s)
		\and
		\inferrule[Equality resolution]
		{
			u \neq v \lor R
		}
		{
			\sigma(R)
		}
		\sigma = \mgu(u, v)
		\and
		\inferrule[Equality factoring]
		{
			s = t \lor u = v \lor R
		}
		{
			\sigma(t \neq v \lor u = v \lor R)
		}
		\sigma = \mgu(s, u)
	\end{mathpar}
	\caption{Selected \vampire inferences. The $\mgu(x,y)$ stands for \emph{the most general unifier} of $x$ and $y$.}
	\label{fig:inferences}
\end{figure}

We now outline the inferences we will be translating for the unfamiliar: refer to \emph{First-Order Theorem Proving and \vampire}~\cite{vampire} for more details.
First, the core rules:
\begin{description}
\item[Resolution] is the propositional resolution rule lifted to first-order clauses. The \emph{most general unifier} can be seen as taking those instances of the premises for which $P$ and $Q$ are identical atoms.
\item[Factoring] merges literals in \emph{instances} of a clause where they are identical. This can be seen as removing duplicate literals in a propositional  clause.
\item[Paramodulation] might be glossed as ``lazy conditional rewriting combined with instantiation''~\cite{E}. At the ground level, it performs case analysis: either $C$ holds, or $l = r$ does. If $l = r$, then we can rewrite $L[s] \vee E$ when $l$ is the same as $s$ under substitution.
\item[Equality Resolution] deletes literals of the form $u \neq v$ when $u$ and $v$ are the same under substitution.
\item[Equality Factoring] is largely a technicality required for completeness. It occurs in proofs very rarely, and we have not yet implemented it.
\end{description}
All of these have various conditions attached in order to reduce the search space.
However, these are extra-logical and not important for verifying the proof, once produced.

There are also various special cases.
\emph{Subsumption resolution}\footnote{Aka \emph{contextual literal cutting}.} is a special case of resolution where the conclusion subsumes one of the premises.
\emph{Duplicate literal removal} is a special case of factoring where the two literals are already identical.
\emph{Demodulation} is a special case of paramodulation where the clause with the rewriting equation has no other literals.
\emph{Definition unfolding} can also be seen as paramodulation, but is done in preprocessing, ahead of time. The inference (true to the name) unfolds definitions given as equalities, by rewriting terms in other clauses.
\vampire also implements \emph{equality resolution with deletion}, which is merely equality resolution performed in preprocessing.
\emph{Trivial equality removal} is also a special case of equality resolution, where the two sides of the equation are already identical.

The default mode of \vampire also incorporates \avatar~\cite{avatar}, a technique that splits certain clauses and offloads the resulting disjunctive structure to a SAT solver.
It is largely orthogonal to the core calculus and is therefore described separately in~\Cref{sec:avatar}.

\section{Scope and Trust}
Here we precisely state the nature of our work, and what must be trusted, in order to avoid confusion.
If \vampire succeeds it will produce a proof by refutation, which is traditionally reported as a human-readable proof following \vampire's internal calculus closely.
Our new \dedukti output mode instead produces a single \dedukti script containing:
\begin{enumerate}
\item\label{axiomatisation} The standard encoding of first-order logic.
\item Some limited shorthand notation (Section~\ref{sec:bridging}).
\item\label{signature} Type declarations of user-declared and \vampire-{in\-tro\-du\-ced} symbols.
\item\label{reproduction} A reproduction of the required axioms (negated conjecture), as parsed.
\item A step-by-step derivation of falsum, following \vampire's inferences.
\end{enumerate}
\dedukti can then check the proof script, and, if all goes well, indicate success.
In order to trust the proof, the user must then trust \dedukti, inspect to their satisfaction the axiomatisation (\ref{axiomatisation}) and reproduction of the input (\ref{reproduction}), and check that no further axioms or rewrites are introduced in the proof script.\footnote{Although, of course, our implementation does not do this.}

Those \vampire inferences resulting from running \vampire in the default mode (i.e. the inferences presented in~\Cref{sec:vampire-calculus}) are implemented, except for the rarely-used equality factoring and steps required to bring the input into normal form.
Problems that are already given in clause normal form --- which is already sufficiently expressive for many applications --- are therefore highly likely to be fully checked.
Inference steps that are not yet supported (see~\Cref{sec:future-work}) are encoded as \texttt{sorry} axioms, which produce a warning during proof checking and must also be trusted.
Failure to pass \dedukti indicates that there is a bug in \vampire, either because \vampire's internal proof is unsound, or because the \dedukti script was generated incorrectly.

The \dedukti script could in principle be translated to an interactive theorem prover.
There, \ref{axiomatisation} and \ref{reproduction} will be checked automatically, and remaining gaps in the proof presumably appear as new subgoals for the user to dispatch manually.
Note that as we \emph{encode} first-order logic in \dedukti (Section~\ref{sec:encoding}), the proof script shows that ``the embedding of falsum can be derived from the embedding of the axioms and the negatated conjecture''. It does \emph{not} show that ``the conjecture follows from the axioms''.
Some translation effort would therefore still be required to extract the embedded proof into the interactive theorem prover's logic.

\section{Bridging Theory and Practice}
\label{sec:bridging}
Naturally, the hardest part of implementing \dedukti output for \vampire is the step-by-step derivation.
We must construct for each step of a given \vampire proof a term $D$ of conclusion type, given constants $C_i$ of premise types already defined.
We are greatly indebted to Guillaume Burel for his schematic $\lambda\Pi$ terms~\cite{burel} covering the standard inferences of the superposition calculus.
We use these almost verbatim to translate core \vampire inferences, with a few important differences that we will now illustrate with an example.

Recall the paramodulation inference rule from~\Cref{sec:vampire-calculus}:
\begin{equation}
\label{eq:superposition-rule}
\inferrule
	{l = r \lor C \\ L[s] \lor E}
	{\sigma(L[r] \lor C \lor E)}
\end{equation}
where $\sigma$ is the \emph{most general unifier} of $l$ and $s$.
Now a concrete instance:
\begin{equation}
	\label{eq:superposition-example}
	\inferrule
	{P(x) \lor f(c,x,z) = g(x) \lor x \neq c \\ Q(y) \lor f(y,d,w) \neq e \lor R(f(c,d,w))}
	{P(d) \lor d \neq c \lor Q(c) \lor g(d) \neq e \lor R(f(c,d,w))}
\end{equation}
In this instance the literals involved are $f(c,x,z) = g(x)$ and $f(y,d,w) \neq e$, with most general unifier $\sigma = \{x \mapsto d, y \mapsto c, z \mapsto w\}$.
Given a signature containing
\iftrue
\begin{equation*}
	\begin{aligned}
		&c : \El{\iota}\\
		&d : \El{\iota}\\
		&e : \El{\iota}\\
	\end{aligned}
	\qquad
	\begin{aligned}
		&f : \El{\iota} \to \El{\iota} \to \El{\iota} \to \El{\iota}\\
		&g : \El{\iota} \to \El{\iota}\\
	\end{aligned}
	\qquad
	\begin{aligned}
		&P : \El{\iota} \to \mathsf{Prop}\\
		&Q : \El{\iota} \to \mathsf{Prop}\\
		&R : \El{\iota} \to \mathsf{Prop}\\
	\end{aligned}
\end{equation*}
\fi
the relevant clause types are
\begin{align*}
&C_1: \Pi x, z : \El{\iota}.~&&\shallow{P(x)} \to  \shallow{f(c,x,z) = g(x)} \to \shallow{x \neq c} \\ 
& &&\to \prf{\false} \\
&C_2: \Pi y, w : \El{\iota}.~&&\shallow{Q(y)} \to  \shallow{f(y,d,w) \neq e} \\
& &&\to \shallow{R(f(c,d,w))} \to \prf{\false}\\
&D : \Pi w : \El{\iota}.~&&\shallow{P(d)} \to  \shallow{d \neq c} \to  \shallow{Q(c)} \\ 
& && \to  \shallow{g(d) \neq e} \to  \shallow{R(f(c,d,w))} \to \prf{\false}
\end{align*}
where $C_1$ and $C_2$ are already shown and $D$ must be proven.
We combine Burel's \emph{instantiation} and \emph{identical superposition} into one step and produce $D$ via
\begin{align*}
& D \hookrightarrow &&\lambda w : \El{\iota}.\\
& &&\lambda \ell_1 : \shallow{P(d)}.~\lambda \ell_2 : \shallow{d \neq c}.\\
& &&\lambda \ell_3 :  \shallow{Q(c)}.~\lambda \ell_4 : \shallow{g(d) \neq e}.~\lambda \ell_5 : \shallow{R(f(c,d,w))}.\\
& &&C_2~c~w~\ell_3~(\lambda q : \prf{\deep{f(c,d,w) \neq e}}.\\
& &&\quad C_1~d~w~\ell_1~(\lambda r : \prf{\deep{f(c, d, w) = g(d)}}.\\
& &&\quad\quad\ell_4~(r~(\lambda z : \El{\iota}.~|z \neq e|)~q))~\ell_2)~\ell_5
\end{align*}
This can be hard to read at first sight.
First, we introduce variables $\lambda w$ and literals $\lambda \ell_i$ to match the type of $D$, and now we must provide a term of type $\prf{\false}$.
First we apply a particular instance $C_2~c~w$ of the main premise computed from the unifier, and use the bound literal $\ell_3$ for its first literal.
When the (instantiated) rewritten literal $f(c, d, w) \neq e$ is encountered, we defer producing a contradiction and instead bind $q$ for later use, continuing with $C_1~d~w$.
When the (instantiated) rewriting literal $f(c, d, w) = g(d)$ is encountered, we again defer and bind $r$.
To produce falsum, we apply $\ell_4$ to an expression representing rewriting $q$ using $r$ in ``context'' $\_ \neq e$.
Finally, $\ell_2$ and $\ell_5$ finish the proof.

\subsection{Calculus Tweaks}
\vampire does in general implement the ``textbook'' superposition calculus, but it also modifies some inference rules and implements extra rules for the sake of efficiency~\cite{vampire}.
Rules which delete clauses altogether do not need to be considered, as they will not appear in the final proof.
Rules which simplify clauses are typically special cases of existing rules (as explained in~\cref{sec:vampire-calculus})
and therefore did not require inventing new $\lambda\Pi$ terms.

Modifications to the calculus can be more tricky.
The \emph{simultaneous} variant of the paramodulation rule rewrites the entire right-hand side, not just the target literal.
In our example, $R(f(c, d, w))$ is also rewritten:
\begin{align*}
	D' :~&\shallow{P(d)} \to  \shallow{d \neq c} \to \shallow{Q(c)} \to \shallow{g(d) \neq e} \\
	& \to \shallow{R(g(d))} \to \prf{\false}
\end{align*}
so the derivation must also construct the rewrite term for the last argument $\ell_5$:
\begin{align*}
& D' \hookrightarrow &&\lambda \ell_1 : \shallow{P(d)}.~\lambda \ell_2 : \shallow{d \neq c}.\\
& &&\lambda \ell_3 : \shallow{Q(c)}.~\lambda \ell_4 : \shallow{g(d) \neq e}.~\lambda \ell_5 : \shallow{R(g(d))}.\\
& &&C_2~c~\inhabit{\iota}~\ell_3~\\
	& &&\quad(\lambda q : \prf{\deep{f(c,d,\inhabit{\iota}) \neq e}}.~C_1~d~\inhabit{\iota}~\ell_1\\
	& &&\quad\quad(\lambda r : \prf{\deep{f(c,d,\inhabit{\iota}) = g(d)}}. \\
	& &&\quad\quad\quad\quad ~\ell_4~(r~(\lambda z : \El{\iota}.~\deep{z \neq e})~q))~\ell_2)\\
	& &&\quad(\lambda q : \prf{\deep{R(f(c,d,\inhabit{\iota}))}}.~C_1~d~\inhabit{\iota}~\ell_1 \\
	& &&\quad\quad(\lambda r : \prf{\deep{f(c,d,\inhabit{\iota}) = g(d)}}.~\ell_5~(r~R~q))~\ell_2)\;
\end{align*}

\subsection{Domains are Inhabited}
\label{sec:inhabit}
You may have noticed that in the above term the variable $w$ disappears from the conclusion, which becomes a ground clause.
However, the derivation needs to provide explicit terms to instantiate $z$ and $w$ in $C_1$ and $C_2$ respectively, so we must explicitly use the fact that $\iota$ is inhabited.
This assumption is usually left tacit in automated theorem proving, but it is there: otherwise one could not resolve $\forall x : \iota.~P(x)$ and $\forall x : \iota.~\lnot P(x)$ to obtain a contradiction as we would not know that there is at least one element of $\iota$.
For this, we use an the explicit ``inhabit'' term $\star$ defined in~\Cref{eq:inhabit-def}.

\subsection{Equality is Symmetric}
\vampire treats equality as symmetric internally, and therefore normalises the orientation of equations.
\dedukti does not treat $\mathsf{eq}$ as symmetric, as it is a user-provided function.
These innocent implementation details cause us significant headache, as whenever any \vampire equation is transformed (typically by rewriting or substitution), it may change its orientation.
It is possible to show symmetry of $\mathsf{eq}$ directly in \dedukti, but it is somewhat verbose to do it inline wherever required.
We therefore use \dedukti's ability to safely introduce definitions to provide helpful commutativity lemmas.
If in our example, the conclusion swaps the orientation of the second literal $c \neq d$, the derivation would need to detect this and tag $\ell_2$ with $(\mathsf{comml\_not} \; c \; d \; \ell_2)$, where $\mathsf{comml\_not}$ is a shorthand for the commutativity lemma with type declaration
\begin{equation*}
	\mathsf{comml\_not} : \Pi x,y : \El~\iota \to \shallow{\lnot \mathsf{eq}~x~y} \to \shallow{\lnot \mathsf{eq}~y~x}.
\end{equation*}

\subsection{Many-Sorted and Polymorphic Logics}
\label{sec:polymorphism}
\vampire has over time grown the ability to have terms of more than one sort~\cite{tff0} and later to support rank-1 polymorphic types~\cite{tff1}, in response to input languages~\cite{tptp-languages,smtlib} and internal requirements~\cite{polymorphic-vampire}.
As \dedukti has a rich type system, it is quite easy to embed these relatively simple systems.
First, each item in the encoding that assumes $\iota$ (such as \textsf{eq}) is instead given a type parameter $\alpha$ to be instantiated later.
Shorthands are similarly adapted.
This already allows a straightforward, if somewhat tedious, implementation of many-sorted logic.

Polymorphism requires more care.
It is not possible to re-use the standard encoding of $\forall$, i.e. $\mathsf{forall} : \Pi \alpha : \mathsf{Set}.~(\El \alpha \to \mathsf{Prop}) \to \mathsf{Prop}$, as $\mathsf{forall}~\mathsf{Set}$ does not type-check.
Instead, we introduce a new binder (as a \dedukti axiom) for universal quantification over sorts, which may be familiar as System F's $\Lambda$.

Implementation can also be confusing.
Consider the clause $P(c) \lor x \neq x$, which \vampire simplifies to $P(c)$.
In a polymorphic setting, this represents the formula $\Lambda \alpha.~\forall x : \alpha.~P(c) \lor x \neq x$.
The sort variable $\alpha$ must be detected and instantiated (in the same way as term variables in Section~\ref{sec:inhabit}) to a don't-care sort in order to apply the simplification in \dedukti: we simply chose $\iota$, but any sort will do.

\subsection{Implementation}
\label{sec:implementation}
The implementation is available,\footnote{
	\url{https://github.com/vprover/vampire}, branch \texttt{dedukti2}
	} but not yet integrated with mainline \vampire.
The flag \texttt{-p dedukti} causes \vampire to produce a \dedukti script instead of its default human-readable proof.
When an inference is not yet supported (Section~\ref{sec:future-work}), the script produces a warning ``sorry'' message, asserts without proof that premises imply the conclusion, and then proves the conclusion from the premises via the assertion.
This is moderately better than just asserting the conclusion, as it ensures premises were consistently handled.

The new output mode requires the additional flag\\ 
\verb|--proof_extra full|.
When enabled, \vampire stores additional information during proof search.
In order to avoid a performance hit we store the minimum possible information and recompute the details during proof printing.
For instance, we store only the participating literals in resolution steps, and can then re-unify them to recompute the required substitutions, the orientation of equations before and after substitution, the variable renaming employed, and so on.
Typical inferences need only a few extra words of memory per clause.

\section{\avatar}
\label{sec:avatar}
\avatar~\cite{avatar} is a distinctive feature of \vampire that greatly improves the efficiency of first-order reasoning in many cases by \emph{splitting} certain clauses and offloading the resulting disjunctive structure to a SAT solver.
For our purposes, this means that we introduce \emph{propositional labels} representing a sub-clause, and must process the following inferences:
\begin{description}
\item[definition] \avatar may introduce a fresh definition for a sub-clause, such as
\begin{align*}
\spl{1} &\equiv \forall xy.~P(x, f(y)) \lor \lnot Q(y)\\
\spl{2} &\equiv \forall z.~c = z
\end{align*}
\item[split] These definitions are used in order to split clauses into variable-disjoint \emph{components}, deriving a SAT clause:
\begin{equation}
\label{eq:split}
\inferrule
	{\lnot Q(z) \lor c = y \lor P(x, f(z))}
	{\spl{1} \lor \spl{2}}
\end{equation}
Note that definitions can be used modulo variable renaming and the order of literals in a clause.

\item[component] Components are injected into the search space as \avatar clauses:
\begin{align*}
P(x, f(y)) \lor \lnot Q(y) &\leftarrow \spl{1}\\
c = z &\leftarrow \spl{2}
\end{align*}
where, for instance, the second clause should be read ``$c = z$ if $\spl{2}$''.
\item[\avatar clauses] All existing inferences must now work conditionally on \avatar splits, taking the union of the parents' \emph{split sets} into the conclusion.
% This is straightforward: each inference must pass extra split parameters to premises as required, but it also \emph{receives} these as parameters itself.
\item[contradiction] It is now possible to derive a contradiction \emph{conditionally} on some \avatar splits, which again derives a SAT clause:
$$
\inferrule
	{\false \leftarrow \spl{3} \land \lnot \spl{5}}
	{\lnot \spl{3} \lor \spl{5}}
$$
\item[refutation] The SAT solver reports that the set of SAT clauses derived is unsatisfiable, and \vampire's proof is finished. At this point \vampire usually reports a single inference deriving falsum using a minimised set of SAT premises, the dreaded \texttt{avatar\_sat\_refutation}. However, \dedukti demands more detailed proofs and so we must extract a proof from the SAT solver.
\end{description}

\subsection{Encoding \avatar Inferences}
A general schema of these inferences as \dedukti terms can be found in~Appendix~\ref{sec:app-avatar}, but it may be hard to understand.
Instead, we rely on the suitably-complex example above to communicate the idea.
An \avatar clause $C \leftarrow \spl{i} \land \lnot\spl{j}$ is represented $\shallow{\lnot \spl{i}} \to \shallow{\lnot\lnot \spl{j}} \to \mathcal{C}$, where $\mathcal{C}$ is the usual representation of $C$ (\ref{eq:clause}).

\begin{description}
\item[definition] We use \dedukti's support for introducing definitions to introduce \avatar definitions without the possibility of unsoundness.
An \avatar definition for a subclause is defined such that $\prf{\spl{i}}$ rewrites to the representation of the subclause (\ref{eq:clause}).
For instance, the definition of $\spl{1}$ is
\begin{equation}
	\label{eq:avatar-def}
	\begin{aligned}
		& \spl{1} : \mathsf{Prop} \\
		& \spl{1} \hookrightarrow \forall x,y : \El{\iota}.~|\lnot P(x, f(y))| \implies |\lnot \lnot Q(y)| \implies \false
	\end{aligned}
\end{equation}

\item[split] Suppose we are preprocessing inference~\eqref{eq:split}, which is sufficiently complex to illustrate the main challenges.
We have the premise
\begin{align*}
C :~&\Pi x, y, z: \El{\iota}.&&~\shallow{\lnot Q(z)} \to \shallow{c = y} \to \shallow{P(x, f(z))} \\ 
& &&\to \prf{\false}
\end{align*}
and must produce a term for:
\begin{align*}
& D : &&\shallow{\spl{1}} \to \shallow{\spl{2}} \to \prf{\false} \\
& D \hookrightarrow &&\lambda s_1: \shallow{\spl{1}}.~\lambda s_2: \shallow{\spl{2}}.\\
	& && s_1~(\lambda x, z : \El \iota.~\lambda \ell_3 : \shallow{P(x, f(x))}.~\lambda \ell_1 : \shallow{\lnot Q(x)}.\\
	& && s_2~(\lambda y : \El \iota.~\lambda \ell_2 : \shallow{c = y}.\\
	& &&\quad C~x~y~z~\ell_1~\ell_2~\ell_3))
\end{align*}
In essence, the split definitions are ``unpacked'' with suitable variable renaming, with resulting variables and literals applied to the premise.
More than two splits generalises naturally, as does using a definition more than once.
\item[component] Morally this term is the identity function, but due to the extra double negation present in the encoding\footnote{You may be tempted to try to eliminate this wart in your own implementation: we ran into trouble when splitting a clause that is already conditional on other splits.} it must be re-packaged somewhat.
For 
\begin{align*}
D' :~&\shallow{\lnot \spl{1}} \to \Pi x, y : \El~\iota.~\shallow{P(x, f(y))} \\ 
     &\to \shallow{\lnot Q(x)} \to \prf{\false},
\end{align*}
the component clause for $\spl{1}$, we do this with
\begin{align*}
	&D' \hookrightarrow &&\lambda \mathsf{nsp}_1 : \shallow{\lnot \spl{1}}.\\
	& &&\lambda x, y: \El~\iota.~\lambda \ell_1 : \shallow{P(x, f(y))}.~\lambda \ell_2 : \shallow{\lnot Q(x)}.\\
	& &&\mathsf{nsp}_1~(\lambda \mathsf{psp}_1 : \shallow{\spl{1}}.~\mathsf{psp}_1~x~y~\ell_1~\ell_2)
\end{align*}
\item[\avatar clauses]
Each conditional inference is now tag\-ged with the \emph{split sets} that need to be taken into account while making the derivation.
Specifically, we need to bind them and then apply them to the parent clauses accordingly.
For instance suppose that we also have $\forall x.~Q(x)$, which translates to $E : \Pi x : \El~\iota.~\shallow{Q(x)}$.
Then we can derive $\forall x,y. P(x, f(y)) \leftarrow \spl{1}$ by resolution:
\begin{align*}
	&D'' : &&\shallow{\lnot \spl{1}} \to \Pi x,y: \El~\iota.~\shallow{P(x, f(y))} \to \prf{\false} \\
	&D'' \hookrightarrow &&\lambda \ell_1 : \shallow{\lnot \spl{1}}.~\lambda x,y : \El~\iota.~\lambda \ell_2 : \shallow{P(x, f(y))}.\\
	& &&D'~\ell_1~x~y~\ell_2~(\mathsf{tnp}: \shallow{Q(x)}.~E~x~\mathsf{tnp})
\end{align*}
Note that the variable $\ell_1$ pertaining to the split $\spl{1}$ has to be bound first and then also applied to the parent clause~$D'$ as the first argument.
\item[contradiction] This is just the premise term.
\end{description}

\subsection{Encoding AVATAR SAT proofs}
Recall that at the end of an AVATAR-assisted proof, the underlying SAT solver shows that all cases have been dispatched by showing unsatisfiability of a set of SAT clauses.
In \dedukti terms, this means that we must derive $\prf \false$ using a given set of propositional clauses.

\vampire uses the SAT solvers \minisat~\cite{minisat} and \cadical~\cite{cadical} internally for various purposes~\cite{uses-of-sat-solvers}, but we used \cadical for reconstructing AVATAR proofs.
\cadical can be configured to output a DRAT~\cite{drat} proof to disk during solving.
This initially appeared challenging as RAT inferences are not deductively valid, but merely preserve satisfiability.
Algorithms exist~\cite{rat2rup} to translate this difficulty away, but in conversation with \cadical developers\footnote{We thank
% Mathias Fleury, Katalin Fazekas and Florian Pollitt 
CaDiCal developers for many helpful discussions about \cadical and \vampire.} it emerged that at present (as of \cadical 2.1.3), only RUP inferences are emitted in the DRAT format.
We mention in passing the \texttt{lrat2dk}~\cite{lrat2dk} tool: this may be helpful if true RAT inferences are needed in future, although it would require producing LRAT~\cite{lrat} proofs from \cadical's DRAT~\cite{cadical-lrat}.

Reverse Unit Propagation (RUP) inferences~\cite{rup} can be seen as a highly-compressed form of multiple resolution steps.
A clause $C$ is a valid RUP inference from premises $\mathcal{D}$ if adding $\lnot C$ to $\mathcal{D}$ and then performing \emph{unit propagation}~\cite{unit-propagation} results in a contradiction.
Fortunately, the process of showing that $C$ is RUP translates very naturally into \dedukti.
First, negating $C$ and propagating the resulting unit literals is the same as introducing the encoded literals via $\lambda$-abstraction.
Then, for any clause $D \in \mathcal{D}$ which unit propagates a literal $\ell$, all other literal holes in $D$ can be dispatched with already-available unit literal terms, while the $\ell$-hole is filled with a ``continuation'' $\lambda$-expression binding $\ell$ as a unit literal.
Eventually some clause is a contradiction in a valid RUP step.
For example, consider the SAT clauses (taken from a \vampire proof of the TPTP problem \texttt{SYN011-1}):
\begin{equation}
\label{sat:432}
4 \vee 3 \vee 2
\end{equation}
\begin{equation}
\label{sat:14}
\lnot 1 \vee \lnot 4
\end{equation}
\begin{equation}
\label{sat:61}
\lnot 6 \vee 1
\end{equation}
\begin{equation}
\label{sat:356}
3 \vee 5 \vee 6
\end{equation}
\begin{equation}
\label{sat:3}
\lnot 3
\end{equation}
\begin{equation}
\label{sat:21}
\lnot 2 \vee \lnot 1
\end{equation}
The unit clause $5$ is RUP from these clauses.
To see this, start by propagating $\lnot 5$.
Clause~\ref{sat:3} also immediately propagates $\lnot 3$.
Using these two literals clause~\ref{sat:356} propagates $6$,
then clause~\ref{sat:61} propagates $1$,
clause~\ref{sat:14} propagates $\lnot 4$ and clause~\ref{sat:432} propagates $2$,
but now clause~\ref{sat:21} produces a contradiction.

The \dedukti term for this RUP inference is
\begin{align*}
C &: \shallow{\spl{5}} \to \prf{\false} \\
C &\hookrightarrow \lambda \ell_{\lnot 5} : \prf{\lnot \spl 5}.\\
&D_{\ref{sat:3}}~(\lambda \ell_{\lnot 3} : \prf{\lnot \spl 3}.\\
&D_{\ref{sat:356}}~\ell_{\lnot 3}~\ell_{\lnot 5}~(\lambda \ell_6 : \prf{\spl 6}.\\
&D_{\ref{sat:61}}~(\lambda \ell_{\lnot 6}.~\ell_{\lnot 6}~\ell_6)~(\lambda \ell_1: \prf{\spl 1}.\\
&D_{\ref{sat:14}}~(\lambda \ell_{\lnot 1}.~\ell_{\lnot 1}~\ell_1)~(\lambda \ell_{\lnot 4}: \prf{\lnot \spl 4}.\\
&D_{\ref{sat:432}}~\ell_4~\ell_3~(\lambda \ell_2: \prf{\spl 2}.\\
&D_{\ref{sat:21}}~(\lambda \ell_{\lnot 2}.~\ell_{\lnot 2}~\ell_2)~(\lambda \ell_{\lnot 1}.~\ell_{\lnot 1}~\ell_1))))))
\end{align*}
The example has been contrived such that all clauses propagate their last literal for the sake of readability.
This not the case in practice, of course, but is easily worked around.

\section{Experiments}
\label{sec:experiments}
\begin{figure}[t]
    \centering
    \includegraphics[scale=0.8]{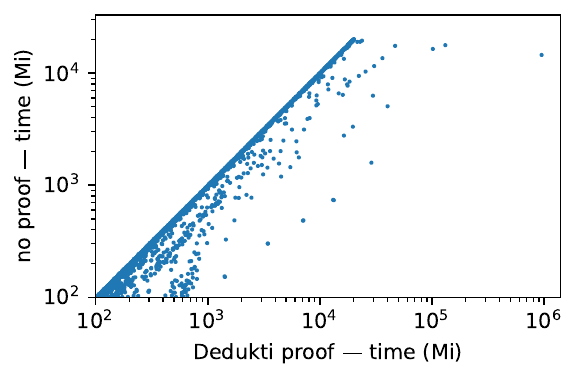}
    \caption{Scatter plot showing the overhead of \dedukti proof generation.}
    \label{fig:scatter}
\end{figure}
In order to test our implementation we ran \vampire on TPTP \cite{Sut24} v9.1.0 problems in the CNF, FOF, TF0 and TF1 fragments.
% Michael's incantation: grep -rm1 SPC Problems | egrep '(FOF|CNF|TF0|TF1)_(UNS|THM|CAX)' | grep -v 'ARI$' | cut -d: -f1 | grep -v .rm > dk-problems
Satisfiable problems and those containing arithmetic were excluded.
\vampire ran using its default strategy (which involves AVATAR), except for replacing the non-{de\-ter\-mi\-nis\-tic} LRS saturation algorithm \cite{DBLP:journals/jsc/RiazanovV03} with its simpler stable variant.
Under an instruction limit of \SI{20000}{Mi} (roughly $\SI{8}{\second}$ on our machine), % I switched to "Intel(R) Xeon(R) Gold 6140 CPU @ 2.30GHz" (from "AMD EPYC 7513 32-Core Processor")
\vampire was able to prove the same set of \num{8178} problems both with proof production disabled and when recording the necessary information to produce detailed \dedukti proofs.
(The average overhead of recording the extra information was $\SI{0.6}{\percent}$.)
The scatter plot in Fig.~\ref{fig:scatter} compares the instructions needed to solve a problem with the instructions needed to solve a problem and additionally generate a \dedukti proof and print it to a file.\footnote{Shown for the \num{8178} problems solved within the instruction limit, but not imposing that limit during proof printing anymore.}
Although on average proof generation only adds \SI{28.0}{\percent} on top of proof search,
the plot shows examples where the overhead was up to 10-fold and around 65-fold for one unfortunate problem.\footnote{This was the problem
\texttt{HWV057+1}, a bounded model checking example translated to first-order logic from QBF \cite{DBLP:conf/cade/SeidlLB12}, featuring exceptionally high Skolem arities. 
The corresponding proof is the largest generated with \SI{1.5}{\gibi\byte}.}
% HWV057+1 948464 14509 65.3707354056103
% (we are lucky not to solve "HWV058+1" or "HWV059+1" anymore, in the described setting)
\begin{figure}[t]
    \centering
    \includegraphics[scale=0.8]{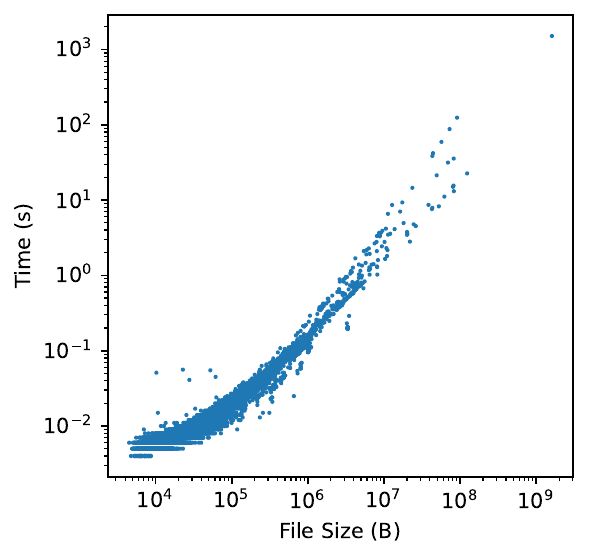}
    \hfill
    \includegraphics[scale=0.8]{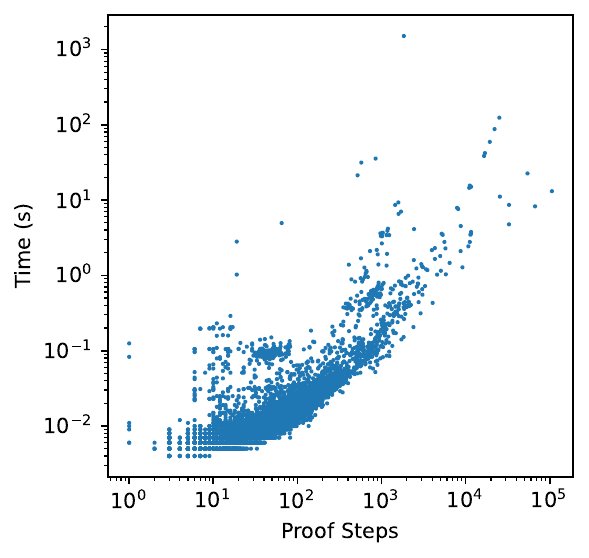} % trim = 10mm 0mm 0mm 0mm, clip,
    \caption{Two scatter plots highlighting \dedukti proof checking time on our proofs.}
    \label{fig:two_plots}
\end{figure}

All proofs were successfully checked by \dedukti{} v2.7. % \footnote{The proofs as well as support scripts are included as part of the supplementary materials.}
% The proof generated from \texttt{SYN986+1.004} caused \dedukti's stack to overflow, but by configuring an unlimited system stack we were able to check it. (not sure about this anymore, since I had `ulimit -s unlimited` already in.
The median time to check a proof was \SI{0.01}{\second}, the mean \SI{0.3}{\second}, and the largest file took \SI{1509}{\second} to check.
%sudamar2@air-04:~/dedukti$ /nfs/sudamar2/TPTP-v9.1.0/dedukti.py
%dk-problems_otter_i20000_dedukti/Problems_HWV_HWV057+1.p.log (True, 1608376217, 1509.819, 1842)
%dk-problems_otter_i20000_dedukti/Problems_SWV_SWV418-1.300.p.log (True, 44098379, 42.009, 16968)
%dk-problems_otter_i20000_dedukti/Problems_SWV_SWV424-1.450.p.log (True, 91258963, 124.216, 25172)
%dk-problems_otter_i20000_dedukti/Problems_SWV_SWV424-1.300.p.log (True, 43188007, 38.58, 16575)
%dk-problems_otter_i20000_dedukti/Problems_SWV_SWV424-1.350.p.log (True, 56935480, 59.34, 19373)
%dk-problems_otter_i20000_dedukti/Problems_SWV_SWV424-1.400.p.log (True, 72606259, 87.67099999999999, 22098)
%dk-problems_otter_i20000_dedukti/Problems_SYO_SYO594+1.p.log (True, 82459132, 35.64, 850)
%dk-problems_otter_i20000_dedukti/Problems_SYO_SYO592+1.p.log (True, 69190630, 31.53, 575)
%dk-problems_otter_i20000_dedukti/Problems_SYO_SYO591+1.p.log (True, 49319862, 21.365, 519)
%dk-problems_otter_i20000_dedukti/Problems_NUM_NUM375+1.030.p.log (True, 81914823, 15.597, 11228)
%dk-problems_otter_i20000_dedukti/Problems_NUM_NUM376+1.030.p.log (True, 80511654, 15.024, 11616)
%dk-problems_otter_i20000_dedukti/Problems_NUM_NUM377+1.030.p.log (True, 123775266, 22.581, 54445)
%Average time: 0.3021080948887324
%Median time: 0.01
%Max time: 1509.819
Figure~\ref{fig:two_plots} correlates checking time against file size and number of proof steps, showing good asymptotic behaviour in practice, even with very large proofs.

\section{Related Work}
The project \emph{Ekstrakto}~\cite{ekstrakto} is a tool for extracting problems from TPTP library and reconstructing proofs from TSTP trace in $\lambda\Pi$-modulo.
The tool is designed to be as general as possible to cater for many ATPs, so it ignores the references to inference rules in the TSTP trace and attempts to re-prove each deduction step with Zenon Modulo~\cite{zenon-modulo}, iProver Modulo~\cite{burel,burel2020firstorder} and ArchSat~\cite{Bury2018SMTSM}; using only the information about which premises were used. Ekstrakto works well on problems specified in clausal normal form, but already struggles with first-order formulas, failing to handle definitions when applied to \vampire proofs. 

The approach taken for translating (higher-order) proofs~\cite{taprogge-melanie-master-leo} from Leo~III~\cite{leo-steen-phd,leo-journal-SteenB21} employs custom made tactics in \lambdapi to aid with finding the correct substitutions for the inferences reported by the automatic prover, and filling out the missing details (like changing orientation of equalities).

An approach closer to ours is a verifier for proofs written in \emph{Theory-Extensible Sequent Calculus (TESC)}~\cite{baek:TESC}, a proof format for ATPs which allows for proof reconstruction. While the calculus subsumes a large fraction of first-order formulas handled by \vampire, it would need to be extended with \vampire-specific features. Three verifiers are currently implemented for TESC, one of which strengthens our trust by being formalized in the ITP Agda~\cite{agda}.

Similar techniques are used on SMT solvers. A recently proposed proof format Alethe~\cite{alethe} enables additional proof verification, either in~\lambdapi~\cite{alethe-to-dedukti}, or by a separate verifier Carcara~\cite{DBLP:conf/tacas/AndreottiLB23}; imported to Isabelle using the $\mathsf{smt}$ tactic~\cite{smt-to-Isabelle}, or imported to Coq via the SMTCoq plugin~\cite{smtcoq}.

One can also verify the proof steps semantically~\cite{gdv}: rather than reconstructing the inferences syntactically we can check that derived proof steps are logically entailed by the parents, using a trusted system like Otter~\cite{otter} for example. While this approach works reasonably well in practice, it still fails if the trusted system failed to check the otherwise sound entailment.

Some ITPs already use \vampire in their proof automation scripts. Isabelle's \emph{Sledgehammer}~\cite{Sledgehammer} calls \vampire to attempt to prove a lemma and, if successful, then uses the information about which axioms/lemmas were used in the proof to try to internally re-prove the result. Such techniques can be upgraded if \vampire already provides a detailed machine-checkable proof.

\section{Conclusion}
The ATP \vampire was adapted to output proofs in the \dedukti concrete syntax for the $\lambda\Pi$-modulo. Although $\lambda\Pi$ is inherently constructive, we could have asserted classical axioms, but we did not have to. \vampire emits a proof of $\false$ from the axioms and the negated conjecture, which turns out to be entirely constructive: only the last step, from $\false$ to conjecture, is missing. It follows from the double negation translation~\cite{kol25,Godel1933,Gentzen1936,Szabo1971-SZATCP-2,Gentzen1974,kuroda1951,Krivine1990,Friedman78:A-translation} that any classical proof can be transformed to this format (with potentially just one last classical step), and we find it interesting that \vampire will already output this naturally.

\subsection{Future Work}
\label{sec:future-work}
While all of \vampire's core inferences are now implemented (apart from equality factoring, for which the implementation is in progress), it is unfortunate that not all \vampire inferences can be checked yet.
Many clausification inferences can be done easily, although the proof graph will need to be constructed with greater detail than presently in order to be checked.
Skolemisation is known to be difficult to verify, but is at least well-studied~\cite{ekstrakto}.
Theories like arithmetic present a challenge: while it is possible to encode these in \dedukti, it is not very efficient and computing e.g. a multiplication of two large numbers may be very slow. Some better support for theories is now available in the \lambdapi proof assistant's standard library, so a migration to \lambdapi concrete syntax for $\lambda\Pi$-calculus modulo may be required. 
There is also a large space of optional \vampire inferences that must be addressed if competition proofs are to be checked~\cite{casc-j12} in future.

Since~\dedukti strives to support proof interoperability, the hope is that \vampire proofs exported in the \dedukti format can be directly translated to other ITPs, enabling more straightforward use of \vampire as a hammer.

% Generally, while we have shown that many \vampire proofs can be computer checked by \dedukti, with some considerable effort, the surface all possible \vampire inferences is very large and may require a re-think or a change of perspective to cover in full\todo{Is this a fair assessment of the situation?}.
% Anja: I would not say that, it may be true but also it might not.

\begin{acks}
    Anja Petković Komel and Michael Rawson were supported by the ERC Consolidator Grant ARTIST 101002685 and the COST Action CA20111 EuroProofNet.
    Martin Suda was supported by the Ministry of Education, Youth and Sports within the dedicated program ERC CZ under the project POSTMAN no. LL1902.
\end{acks}

\bibliographystyle{ACM-Reference-Format}
\bibliography{cpp26}

\newpage
\appendix

\section{\avatar general schema}
\label{sec:app-avatar}
Here a general schema is given how to translate \avatar inferences into $\lambda\Pi$-modulo proofs. Where applicable we use the example from~Section~\ref{sec:avatar} to illustrate the notation.
Let
$
	L_1 \lor L_2 \lor \ldots \lor L_n
$
be the clause that \avatar splits, and
\begin{equation*}
	C : \Pi x_1, \ldots, x_n : \El{\iota}. \; \shallow{L_1} \to \ldots \to \shallow{L_n} \to \prf{\false}
\end{equation*}
its \dedukti translation.

Let $\{S_i\}_{i\in \{1,2,\ldots,k\}}$ be the partition of the clauses set \\ 
$\{L_i\}_{i \in \{1,2,\ldots,n\}}$ chosen by \avatar. In our example the partition would be $S_1 = \{P(x, f(y)),  \lnot Q(y)\}$ and $S_2 = \{c = z\}$.
For each split $i \in \{1,2,\ldots,k\}$ where $S_i = \{L^i_1, \ldots, L^i_{n_i}\}$ we define a new \dedukti constant
\begin{align*}
	& \spl{i} : \mathsf{Prop} && \\
	& \spl{i} \hookrightarrow \forall y^i_1, \ldots, y^i_{m_i} : &&\El{\iota}. \; \neg L^i_1 \implies \neg L^i_2 \implies \ldots  \\ 
	& &&\implies \neg L^i_{n_i} \implies \false 
\end{align*}
which corresponds to formula~\ref{eq:avatar-def} in our example.
Note that here $L^i_j$ are actual formulas (of type $\mathsf{Prop}$), not the interpreted literals in a clause.

\begin{description}
	\item[split clause]
\end{description}

The parent clause can then be represented by a split clause
\begin{equation*}
	\spl{1} \lor \spl{2} \lor \ldots \lor \spl{k}
\end{equation*}
The derivation of the \avatar split clause will be
\begin{align*}
	& D : \shallow{\spl{1}} \to \shallow{\spl{2}} \to \ldots \to \shallow{\spl{k}} \to \prf{\false} \\
	& D \hookrightarrow
	\lambda s_1 : \shallow{\spl{1}}. \; \ldots \; \lambda s_k : \shallow{\spl{k}}. \;\\
	& \qquad s_1 (\lambda y^1_1, \ldots, y^1_{m_1} : \El{\iota}. \; \lambda \ell^1_1 : \prf{ (\neg L_1^1)}. \; \ldots \; \lambda \ell^1_{n_1} : \prf{(\neg L_{n_1}^1)}. \\
	& \qquad s_2 (\ldots \\
	& \qquad \vdots \\
	& \qquad s_k (\ldots \\
	& \qquad \qquad C \; (\sigma(x_1)) \; \ldots \; (\sigma(x_m)) \; (\sigma(\ell_1)) \; \ldots \; (\sigma(\ell_n)) )) \ldots )
	% & \qquad \qquad \mathsf{parent}\; y^i_1\; \ldots \; y^k_{m_k} \; \ell_1^1 \; \ldots \; \ell_{n_k}^k )) \ldots )
\end{align*}
where $\sigma$ is the substitution of the $x$-variables and $\ell$-variables pertaining to the original term $C$ to the $y$-variables and $\ell$-variables of the corresponding split.
For instance in our example the variable $z$ appears in the split $\spl{1}$ presented by $\sigma(z) = y_1^2$; and the third literal in the parent clause $C$ is $P(x,f(z))$ which appears in the split $\spl{1}$ in the first place, so $\sigma(\ell_3) = \ell_1^1$.

Note that the literal variables $\ell_i^j$ are binding the proof of the negation of the original literals, but since the representation in the clause of the parent $\shallow{L^i_j}$ are also rewritten to $\prf{L_j^i} \to \prf{\false}$ the arguments have matching types.

\begin{description}
	\item[component clause]
\end{description}

The \avatar component clause for the $i$-th split component is morally
\begin{equation*}
	\spl{i} \implies L^i_1 \lor L^i_2 \lor \ldots \lor  L^i_{n_i}
\end{equation*}
which should have a trivial derivation if we unfold the definition of $\spl{i}$.
\begin{equation*}
	(\neg \spl{i}) \lor L^i_1 \lor L^i_2 \lor \ldots \lor  L^i_{n_i}
\end{equation*}
We unfold and re-fold the lambda abstraction as follows.
\begin{align*}
	& D : \shallow{\neg \spl{i}} \to \Pi y^i_1, \ldots, y^i_{m_i} : \El{\iota}. \;  \shallow{L^i_1} \to \ldots \to \shallow{L^i_{n_i}} \to \prf{\false} \\
	& D \hookrightarrow \lambda \mathsf{nsp}_i : \shallow{\neg \spl{i}}. \; \\
	& \qquad \qquad \lambda y^i_1, \ldots, y^i_{m_i} : \El{\iota}.\\
	& \qquad \qquad \lambda \ell^i_1 : \shallow{L_i^1}. \; \ldots \; \lambda \ell^i_{n_i} : \shallow{L_{n_i}^i}. \\
	& \qquad \qquad \mathsf{nsp}_i \; (\lambda \mathsf{psp}_i : \shallow{\lnot \spl{i}}. \; \mathsf{psp}_i \; y^i_1\; \ldots \; y^i_{m_i} \; \ell^i_1 \; \ldots \;  \ell^i_{n_i})
\end{align*}

\section{\dedukti ``prelude'' listing}
The \dedukti prelude that we use for all \vampire proofs is reproduced in Figure~\ref{fig:prelude} below.
Mostly it is standard, but we draw attention to \texttt{inhabit} (\Cref{sec:inhabit}), \texttt{forall\_poly} (\Cref{sec:polymorphism}), the specialisation of \texttt{or} (\Cref{sec:encoding}), the commutativity lemmata \texttt{comm*} and the various \texttt{Prf\_clause} and \texttt{Prf\_av\_clause} shorthands.
\begin{figure*}
\lstset{basicstyle=\tiny\ttfamily}
\begin{lstlisting}
(; Prop ;)
Prop : Type.
def Prf : (Prop -> Type).
true : Prop.
[] Prf true --> (r : Prop -> ((Prf r) -> (Prf r))).
false : Prop.
[] Prf false --> (r : Prop -> (Prf r)).
not : (Prop -> Prop).
[p] Prf (not p) --> ((Prf p) -> (r : Prop -> (Prf r))).
and : (Prop -> (Prop -> Prop)).
[p, q] Prf (and p q) --> (r : Prop -> (((Prf p) -> ((Prf q) -> (Prf r))) -> (Prf r))).
or : (Prop -> (Prop -> Prop)).
[p, q] Prf (or p q) --> (((Prf p) -> (Prf false)) -> (((Prf q) -> (Prf false)) -> (Prf false))).
imp : (Prop -> (Prop -> Prop)).
[p, q] Prf (imp p q) --> ((Prf p) -> (Prf q)).
iff : Prop -> Prop -> Prop.
[p, q] Prf (iff p q) --> (Prf (and (imp p q) (imp q p))).

(; Set ;)
Set : Type.
injective El : (Set -> Type).
iota : Set.
inhabit : A : Set -> El A.

(; Equality ;)
def eq : a : Set -> El a -> El a -> Prop.
[a, x, y] Prf (eq a x y) --> p : (El a -> Prop) -> Prf (p x) -> Prf (p y).
def refl : (a : Set) -> x : (El a) -> Prf (eq a x x).
[a, x] refl a x --> p : ((El a) -> Prop) => t : Prf (p x) => t.
def comm : (a : Set) -> x : (El a) -> y : (El a) -> Prf (eq a x y) -> Prf (eq a y x).
[a, x, y] comm a x y --> e : (Prf (eq a x y)) => p : ((El a) -> Prop) => e (z : (El a) => imp (p z) (p x)) (t : (Prf (p x)) => t).
def comml : (a : Set) -> x : (El a) -> y : (El a) -> (Prf (eq a x y) -> Prf false) -> (Prf (eq a y x) -> Prf false).
[a, x, y] comml a x y --> l : (Prf (eq a x y) -> Prf false) => e : Prf (eq a y x) => l (comm a y x e).
def comml_not : (a : Set) -> x : (El a) -> y : (El a) -> (Prf (not (eq a x y)) -> Prf false) -> (Prf (not (eq a y x)) -> Prf false).
[a, x, y] comml_not a x y --> l : ((Prf (eq a x y) -> Prf false) -> Prf false) => ne : (Prf (eq a y x) -> Prf false) =>
                              l (e : Prf (eq a x y) => ne (comm a x y e)).

(; Quant ;)
forall : (a : Set -> (((El a) -> Prop) -> Prop)).
[a, p] Prf (forall a p) --> (x : (El a) -> (Prf (p x))).
exists : (a : Set -> (((El a) -> Prop) -> Prop)).
[a, p] Prf (exists a p) --> (r : Prop -> ((x : (El a) -> ((Prf (p x)) -> (Prf r))) -> (Prf r))).

(; polymorphic quantifier ;)
forall_poly : (Set -> Prop) -> Prop.
[p] Prf (forall_poly p) --> a : Set -> Prf (p a).

(; Classic ;)
def cPrf : (Prop -> Type) := (p : Prop => (Prf (not (not p)))).
def cand : (Prop -> (Prop -> Prop)) := (p : Prop => (q : Prop => (and (not (not p)) (not (not q))))).
def cor : (Prop -> (Prop -> Prop)) := (p : Prop => (q : Prop => (or (not (not p)) (not (not q))))).
def cimp : (Prop -> (Prop -> Prop)) := (p : Prop => (q : Prop => (imp (not (not p)) (not (not q))))).
def cforall : (a : Set -> (((El a) -> Prop) -> Prop)) := (a : Set => (p : ((El a) -> Prop) => (forall a (x : (El a) => (not (not (p x))))))).
def cexists : (a : Set -> (((El a) -> Prop) -> Prop)) := (a : Set => (p : ((El a) -> Prop) => (exists a (x : (El a) => (not (not (p x))))))).

(; Clauses ;)
def prop_clause : Type.
def ec : prop_clause.
def cons : (Prop -> (prop_clause -> prop_clause)).
def clause : Type.
def cl : (prop_clause -> clause).
def bind : (A : Set -> (((El A) -> clause) -> clause)).
def bind_poly : (Set -> clause) -> clause.
def Prf_prop_clause : (prop_clause -> Type).

[] Prf_prop_clause ec --> (Prf false).
[p, c] Prf_prop_clause (cons p c) --> ((Prf p -> Prf false) -> (Prf_prop_clause c)).
def Prf_clause : (clause -> Type).
[c] Prf_clause (cl c) --> (Prf_prop_clause c).
[A, f] Prf_clause (bind A f) --> (x : (El A) -> (Prf_clause (f x))).
[f] Prf_clause (bind_poly f) --> (alpha : Set -> (Prf_clause (f alpha))).

def av_clause : Type.
def acl : clause -> av_clause.
def if : Prop -> av_clause -> av_clause.
def Prf_av_clause : av_clause -> Type.

[c] Prf_av_clause (acl c) --> Prf_clause c.
[sp, c] Prf_av_clause (if sp c) --> (Prf (not sp) -> Prf false) -> Prf_av_clause c.
\end{lstlisting}
\caption{\dedukti prelude used for all \vampire proofs.}
\label{fig:prelude}
\end{figure*}

\end{document}